\documentclass[twocolumn,showpacs,preprintnumbers,amsmath,amssymb,epsfig]{revtex4}

\usepackage{graphicx}
\usepackage{dcolumn}
\usepackage{bm}
\usepackage{epsf}

\def\Mp{\mbox{${ m _{\rm p} }$\,}}
\def\Mc{\mbox{${ m _{\rm c} }$\,}}
\def\Mtot{\mbox{${ m _{\rm tot} }$\,}}
\def\Msun{\mbox{${ {\rm M} _{\odot} }$\,}}
\def\omegadot{\mbox{${ \dot \omega }$\,}}
\def\Pb{\mbox{${ P _{\rm b} }$\,}}

\def\Pbdot{\mbox{${ \dot P _{\rm b} }$\,}}
\def\Pbdotmeas{\mbox{${ \dot P _{\rm b} ^{\rm meas} }$\,}}
\def\PbdotGR{\mbox{${ \dot P _{\rm b} ^{\rm GR} }$\,}}
\def\Pbdotkin{\mbox{${ \dot P _{\rm b} ^{\rm kin} }$\,}}
\def\PbdotGal{\mbox{${ \dot P _{\rm b} ^{\rm Gal} }$\,}}
\def\VT{\mbox{${ v _{\rm t} }$\,}}
\def\To{\mbox{${ T _{\rm 0} }$\,}}

\def\velu{\mbox{${\rm km \, s^{-1} }$\,}}

\def\alphazerosq{\mbox{${ \alpha _0 ^2 }$\,}}
\def\alphazero{\mbox{${ \alpha _0 }$\,}}
\def\betazero{\mbox{${ \beta _0 }$\,}}

\def\sns{\mbox{$ \varepsilon _{\rm ns} $}}
\def\snsq{\mbox{$ \varepsilon _{\rm ns} ^2 $}}
\def\apsi{\mbox{${ a (\psi) }$\,}}
\def\psisq{\mbox{${ \psi ^2 }$\,}}

\def\la{\mbox{\raisebox{-0.1ex}{$\scriptscriptstyle \stackrel{<}{\sim}$\,}}}

\begin{document}


\preprint{To Appear in Physical Review D}

\title{Gravitational-radiation losses from the pulsar--white-dwarf binary PSR J1141$-$6545}

\author{N. D. Ramesh Bhat, Matthew Bailes, and Joris P. W. Verbiest}
\affiliation{%
Centre for Astrophysics \& Supercomputing, Swinburne University of Technology, Hawthorn, Victoria 3122, Australia.
}%

\date{\today}

\begin{abstract}
Pulsars in close binary orbit around another neutron star or a massive white dwarf make ideal laboratories 
for testing the predictions of gravitational radiation and self-gravitational effects.
We report new timing measurements of the pulsar--white-dwarf binary PSR J1141$-$6545, providing strong 
evidence that such asymmetric systems have gravitational wave losses that are consistent with general 
relativity. 
The orbit is found to be decaying at a rate of $1.04 \pm 0.06$ times the general relativistic prediction and 
the Shapiro delay is consistent with the orbital inclination angle derived from scintillation measurements.
The system provides a unique test-bed for tensor-scalar theories of gravity; our current measurements place 
stringent constraints in the theory space, with a limit of $ \alphazerosq < 2.1 \times 10^{-5} $  for weakly 
non-linear coupling and an asymptotic limit of $ \alphazerosq < 3.4 \times 10^{-6} $ for strongly non-linear
coupling, where \alphazero is the linear coupling strength of matter to an underlying scalar field. This asymptotic
limit is nearly three times smaller than the Cassini bound ($ \alphazerosq \approx 10^{-5} $).
\end{abstract}

\pacs{04.30.Tv,  04.80.Cc,  95.30.Sf,  95.85.Sz}
\maketitle

{\it Introduction.}---Einstein's general theory of relativity (GR) has passed all experimental tests 
so far with complete
success, which makes it one of the most celebrated theories in modern physics \cite{will2006}.
While the most precise tests have been conducted in the weak-field conditions of the solar system
\cite{bertottietal2003,williamsetal2004}, massive and compact astronomical objects such as neutron 
stars and black holes, in particular pulsars in close binary orbits with another neutron star or a 
massive white dwarf (i.e. relativistic binary pulsars), allow testing GR in strong-field conditions 
\cite{tayloretal1992,stairsetal2002,krameretal2006}.
These tests have confirmed GR at an impressive level of better than 1\%
\cite{bertottietal2003,williamsetal2004,tayloretal1992,stairsetal2002,krameretal2006}.

While GR is indeed the most successful theory of gravity, it may conceivably break down 
under extreme strong-field conditions where other theories of gravity may apply. 
Tensor-scalar theories, which invoke a coupling between matter and an underlying 
scalar field (in addition to the standard space-time tensor field), are thought
to be the most natural alternatives to GR \cite{de92,esposito2005}. 
The most well-known in this framework is the Jordan-Brans-Dicke theory,  with a 
linear coupling between matter and scalar field: $ \apsi=\alphazero \psi $,
whilst a more general description involves a two-dimensional parameter space: 
$ \apsi = \alphazero \psi + {1 \over 2} \betazero \psisq $, where $ \psi $
is the strength of the scalar field; $ \apsi $ is the coupling strength between 
the scalar field and matter; \alphazero and \betazero are the coupling parameters. 
The scalar field does not exist in GR and therefore $(\alphazero, \betazero) = (0, 0) $.
Possible deviations from GR are thus constrained by experimentally imposed bounds on 
\alphazero and \betazero. While several tests have been devised for this (see 
\cite{damour2007,esposito2005,stairs2003} for recent reviews), current best limits come 
from timing binary pulsars ($ \betazero > -4.5$) and the Cassini time-delay experiment 
($\alphazerosq < 10^{-5} $; \cite{bertottietal2003}).

The binary-pulsar tests so far 
\cite{tayloretal1992,stairsetal2002,krameretal2006}
have focused on systems that consist of two almost identical
neutron stars (i.e. symmetric systems in which both the objects are of nearly equal mass and size).
PSR J1141$-$6545, on the other hand, comprising a strongly self-gravitating neutron star and a 
(relatively) weakly self-gravitating white dwarf, is a gravitationally asymmetric system and thus 
provides a unique laboratory for testing GR and alternative theories of gravity \cite{esposito2005}.
The asymmetry here is due to the self-gravity ($\varepsilon$) or the compactness of the two bodies
(given by $ \varepsilon = - G M / (R c^2) $, where $M$ and $R$ are mass and radius respectively).
Since $ \varepsilon \sim  - 0.2 $ for a neutron star and $ \sim - 10^{-4} $ for a white dwarf, a neutron star-white 
dwarf binary is a very asymmetric system.

Discovered in 1999, PSR J1141$-$6545 is a relatively young pulsar (characteristic age $\sim$1.4 Myr) spinning at 
a rate of once every 394 ms and is in a 4.74-hr orbit with a moderate eccentricity of $\sim$0.17  
\cite{kaspietal2000}.
Early studies suggested that the pulsar lies at a minimum distance of 3.7 kpc \cite{ordetal2002a} 
and the orbit is inclined at an angle of $ 76^{\circ} \pm 2^{\circ}.5 $ with respect to the plane 
of the sky \cite{ordetal2002b}. 
The orbital period derivative from initial timing analysis was shown to be consistent with GR
at the 25\% level \cite{bailesetal2003} and, more recently, observations over a 6-yr time span 
have unveiled remarkable changes in both the pulse shape and polarisation, which are attributed
to geodetic precession resulting from general relativistic spin-orbit coupling \cite{hotanetal2005}.

In this article, we report measurements of four post-Keplerian parameters for PSR J1141$-$6545:
advance of periastron \omegadot, time dilation and gravitational redshift $\gamma$, the Shapiro 
delay parameter $s$, and the orbital period derivative \Pbdot, that allow us to determine 
the masses of the pulsar and its companion; demonstrate that gravitational wave radiation losses 
are consistent with those predicted by GR; and place stringent limits on tensor-scalar theories 
of gravity.

\smallskip

{\it Timing Observations and Analysis.}---Timing observations of PSR J1141--6545 were undertaken 
using the 64-m Parkes radio telescope in New South Wales, Australia, between 2001 January and 2007 May.
Data were recorded primarily in the 20\,cm radio band, 
while limited data were gathered in the 50\,cm band.
Early observations (2001 to 2002) made use of a multichannel filterbank as the backend,
recording data over a bandwidth of 256 MHz, while data from 2003 onward 
were recorded using the Caltech-Parkes-Swinburne baseband recorder 2,
which coherently 
dedisperses data over two adjacent 64-MHz bands, thus yielding a net bandwidth of 128 MHz.
The data were folded at the predicted topocentric period of the pulsar, where
the adopted integration time varied from 60 seconds (2001 to 2003)
to 300 seconds 
(2004 onward), to account for almost a factor of two 
decrease observed in the pulsar's mean flux density over this period.
The measurement uncertainties in the pulse arrival times have significantly degraded over
this period, as a result of a dramatic increase observed in the pulse width (nearly by a 
factor of two over seven years) probably caused by precession of the pulsar spin axis.

The pulse arrival times were computed by correlating the observed pulse profiles with 
template profiles of high signal-to-noise ratio constructed from long integrations of 
data.  
A total of 12,842 pulse times-of-arrival (TOAs) were measured. 
The remarkable changes observed in the pulse profile over our seven-year long observing 
span, while exciting for studies of geodetic precession and modelling the pulsar 
emission geometry, have complicated our timing analysis.  
In order to minimise the systematics caused by secular profile changes, we adopted a 
strategy which involves the use of a standard (template) profile that is a function 
of time. 
The final TOAs were analysed using the standard timing package {\tt TEMPO},
fitting for the pulsar spin, astrometric and Keplerian parameters, as well as three 
post-Keplerian parameters according to the relativistic and theory-independent timing 
model of Damour and Deruelle (DD) \cite{dd86}.
The final root mean square post-fit residual is 154 $\mu$s.  
The measured pulsar and binary system parameters are listed in Table 1.

\begin{table}
\caption{\label{tab:table}Timing parameters of PSR J1141$-$6545.}
\begin{ruledtabular}
\begin{tabular}{ll}
Spin and astrometric parameters: & \\
Right ascension, $\alpha$ (J2000) \dotfill              &       11$^h$41$^m$07$^s$.0140(2)      \\
Declination, $\delta$ (J2000)   \dotfill & $-$65$^{\circ}$45$^{\prime}$19$^{\prime\prime}$.1131(15) \\
Pulse frequency, $\nu$ ($ {\rm s^{-1}}$) \dotfill       &       2.538729369926(1)       \\
Reference epoch (MJD)    \dotfill               &       51369.852500    \\
Dispersion measure (${\rm pc \, cm ^{-3}}$) \dotfill     &      116.080(1)      \\
First derivative of pulse frequency (${\rm s^{-2}}$) \dotfill &
$-$2.767986(1) $\times 10^{-14}$         \\
 & \\
Keplerian parameters: & \\
Orbital period, \Pb (days) 	\dotfill &	0.1976509593(1) 	\\
Projected semi-major axis, $x$ (s) 	\dotfill &	1.858922(6) 	\\
Orbital eccentricity, $e$               \dotfill & 	0.171884(2) 	\\
Time of periastron passage, $ \To $ (MJD)    \dotfill &	51369.8545515(9) 	\\
Longitude of periastron, $\omega$ ($^{\circ}$) \dotfill &	42.4561(16) 	\\
  & \\
Post-Keplerian parameters: & \\
Advance of periastron, $ \dot \omega $ (${\rm ^{\circ} \, yr^{-1}}$) \dotfill & 5.3096(4) 	\\
Gravitational redshift parameter, $\gamma$ (ms) \dotfill &	 0.000773(11) 	\\
Orbital period derivative, \Pbdot \dotfill & $-$0.403(25) $\times 10^{-12}$ 	\\
\end{tabular}
\end{ruledtabular}
Figures in parentheses represent the nominal 1-$\sigma$ uncertainties
in the least-significant digits quoted.
\end{table} 

\smallskip

{\it Tests of General Relativity.}---In double neutron star systems and in close 
eccentric systems such as PSR J1141$-$6545, the gravitational fields are so strong 
that the application of relativistic gravity becomes essential in timing models. 
For such systems, the observed pulse arrival times are modified by relativistic 
effects, which are potentially measurable through long-term (several year)
timing observations.
These relativistic effects may manifest themselves in various ways; for instance 
a temporal change in the period or orientation of the orbit, or an additional time 
delay (the Shapiro delay) resulting from the curvature of space-time when pulses 
pass near the massive companion.
These effects can be modelled in terms of the post-Keplerian (PK) parameters, 
which are essentially some phenomenological corrections and additions to the 
Keplerian orbital parameters \cite{dd86}.
These PK parameters have different dependencies in different theories of gravity, 
and thus facilitate important tests of theories of gravity \cite{will2006,dt92}.
In any theory of gravity, the PK parameters can be expressed as functions of the 
pulsar and companion masses and the easily measurable Keplerian parameters.
A binary pulsar requires two PK parameters to completely determine the inclination 
and masses of the system.
Measurement of three or more PK parameters over-determines the system, and thus 
can provide one (or more) test(s) of gravitational theories through self-consistency 
checks.

The orbital period derivative \Pbdot is a crucial parameter as it is still the
only observable that has ever verified the existence of gravitational waves.
Relative to previous timing analysis \cite{bailesetal2003}, 
the measurement precisions on the PK parameters 
have now improved by up to a factor of 4 and, most notably, \Pbdot is determined to be 
$ ( -4.03 \pm 0.25 ) \, \times \, 10^{-13} $.
This measured value (\Pbdotmeas) is a combination of the orbital decay due to
the emission of gravitational radiation (\PbdotGR) and contributions resulting
from both real and apparent accelerations of the binary system along the line
of sight; i.e.,
\begin{equation}
\Pbdotmeas = \PbdotGR + \Pbdotkin + \PbdotGal
\end{equation}
where ``Gal'' and ``kin'' refer to the Galactic and kinematic contributions
respectively \cite{dt91}.
The kinematic contribution, known as the Shklovskii effect (an apparent
acceleration of the system due to its space motion, \VT), is given by
$ v _t ^2 / ( c \, d ) $, where $d$ is the pulsar distance and $c$ is the
speed of light.
The Galactic contributions include differential rotation in the plane of the
Galaxy and acceleration in the Galactic gravitational potential.
Given a transverse speed of $ 115 \pm 15 $ \velu deduced from
scintillation observations and a pulsar distance of 3.7 kpc, we
estimate \Pbdotkin = $ (6.71 \pm 1.73) \times 10^{-15} $
and \PbdotGal = $ (-5.05 \pm 0.44) \times 10^{-15} $.
Subtracting these 
from the measured \Pbdot, we obtain an
intrinsic value 
$ (-4.01 \pm 0.25) \times 10^{-13} $ for \PbdotGR.
This is $ 1.04 \pm 0.06 $ times the general relativistic prediction 
and corresponds to a shrinkage of the pulsar's orbit at a rate of 
approximately 2 mm per day.

As the measurement of any two PK parameters allows solving for the two unknown stellar
masses, our measurement of three PK parameters offers an independent test of
GR.
These results can be displayed elegantly in a ``mass-mass'' diagram as shown in Figure 1.
Measurement of the PK parameters gives curves on this diagram that are, in general,
different for different theories of gravity but should intersect at a single point
(i.e. at a pair of mass values) if the theory is valid.
Our results confirm that GR is a correct theory of gravity 
for asymmetric binary systems, thereby extending the range of systems for which GR provides 
an accurate description.

\begin{figure}[b]
\begin{center}
\epsfxsize=86mm \epsfbox{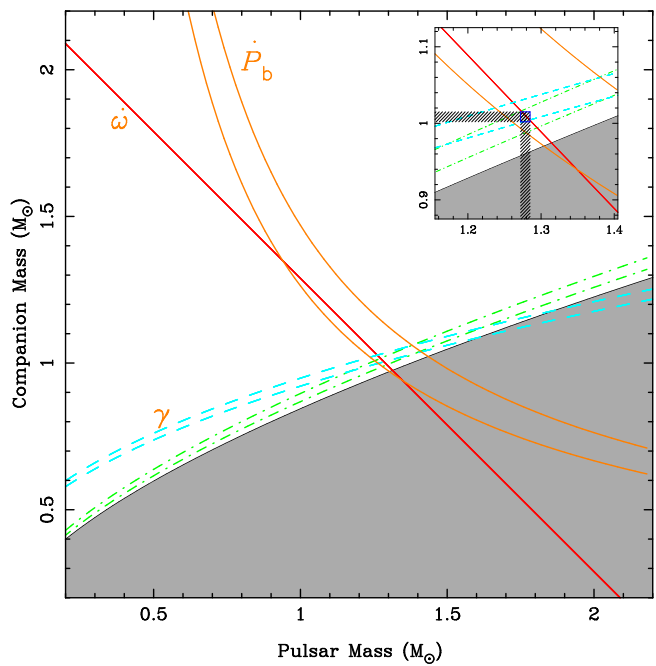}
\caption{Graphical summary of relativistic parameters from timing measurements 
of the PSR J1141$-$6545 binary system. 
The shaded area is excluded from a consideration of Kepler's laws and 
the other constraints are based on the measured PK parameters (shown as 
pairs of lines, with the line separation indicating the measurement
uncertainty) interpreted within the framework of GR.
The systemic mass \Mtot = $2.2892 \pm 0.0003$ \Msun is very accurately 
defined from \omegadot and is shown by the straight solid lines; the 
$\gamma$ term constrains the system to lie between the dashed lines,
and \Pbdot between the pair of curved solid lines.
The $\sin \, i$ measurement from scintillation analysis \cite{ordetal2002b}
is shown as the pair of dot-dashed lines. 
The hatched areas (inset) accentuate the masses allowed within the framework of GR.
}
\label{fig:m1m2}
\end{center}
\vskip-1.0cm
\end{figure}

Given this, it is justifiable to
apply the Damour Deruelle GR (DDGR) formalism (as implemented in {\tt TEMPO}) to
determine the pulsar and companion masses \cite{dd86,dt92}.
This timing model is in the framework of DD but assumes GR to be the true theory of
gravity.
It uses measurements of the PK parameters \omegadot and $\gamma$
to solve for the companion and total masses of the system (\Mc and \Mtot respectively)
and to model the Shapiro delay.
The pulsar mass ($m_{\rm p}$) is then given by $\Mtot-\Mc$.
The best-fit values from such a DDGR fit to our timing 
data are $\Mtot = 2.2892 \pm 0.0003$
\Msun and $ \Mc = 1.02 \pm 0.01 $ \Msun, which implies 
the pulsar mass \Mp = $1.27 \pm 0.01$ \Msun.
These mass estimates
imply an orbital inclination angle ($i$)
of $ 73^{\circ} \pm 1^{\circ}.5 $.
This is in excellent agreement with independent constraints of $ i = 76^{\circ} \pm 2^{\circ}.5 $
derived from the orbital modulation of the pulsar's scintillation velocity \cite{ordetal2002b}.

\begin{figure}
\vskip0.4cm
\begin{center}
{\hskip 0.0cm
\epsfxsize=70mm \epsfbox{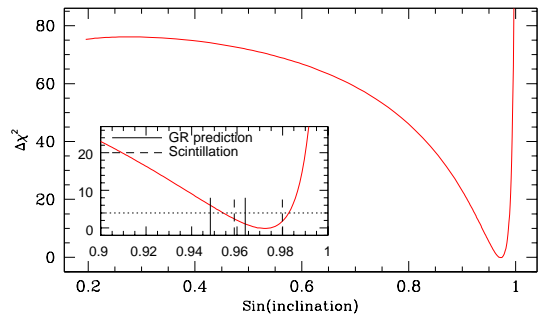}
}
\caption{Current constraints on the orbital inclination angle ($i$) of the PSR J1141$-$6545 system. 
The smooth curve represents values of $ \Delta \chi ^2 $ obtained when sin $i$ was held 
fixed at a range of values from 0.2 to 1 (i.e. $i$ from $10^{\circ}$ to $90^{\circ}$) and 
the companion mass at the value determined by \Mtot = 2.2892 \Msun and the mass function 
($f_p$) = 0.176701 \Msun.
}
\label{fig:sini}
\end{center}
\vskip-0.8cm
\end{figure}

Figure \ref{fig:sini} shows a goodness-of-fit plot for sin $i$ values between 0.2 and 1, in which
the companion mass was set to a value determined by the precisely known systemic mass \Mtot and
the pulsar mass function, $f_p = (\Mc $sin$ i)^3 / m_{\rm tot}^2$ = 0.176701 \Msun.
Such an approach is justifiable as the relativistic \omegadot \,is determined at a very high precision 
and hence traces out a unique locus in the mass-mass diagram, with each point on the locus implying 
a unique inclination angle.  
While the use of the DDGR-derived \Mtot means assumption of GR, the timing model used is DD.
The two-sigma range in the inclination angle ($72^{\circ}.6$ to $79^{\circ}.4$) indicated by this curve is 
consistent with completely independent constraints derived from the scintillation measurements and thus, 
effectively, provides yet another confirmation of GR in this system.

The revised mass estimate of PSR J1141$-$6545 makes it one of the lighter neutron stars
currently known.
Although the pulsar's progenitor was initially the lighter star in the original binary pair,
mass transfer from the white dwarf's progenitor made it large enough to explode as
a supernova even though its companion was not large enough to do so \cite{tauris-sennels2000}.
Being the second star to evolve in the binary, PSR J1141$-$6545 is somewhat similar to pulsar B
in the double pulsar system and the companion to PSR J1756$-$2251 \cite{krameretal2006,faulkneretal2005}. 
It seems probable
therefore, that the progenitors to these pulsars were of lower mass and that the mass
of a neutron star is related to that of the progenitor. Recycled pulsars, or those
with higher eccentricities, appear to be heavier \cite{faulkneretal2005,bailes2007}
either due to mass accretion \cite{jacobyetal2005} or higher initial masses.

\smallskip

{\it Alternative Theories of Gravity.}---Deviations from GR inherent to most alternative 
theories of gravity are best detectable in strong gravitational fields, and hence 
binary-pulsar measurements are indispensable in testing such theories \cite{de96}.
In particular, the gravitationally asymmetric nature of the 
PSR J1141$-$6545 system makes it 
a unique testing ground for tensor-scalar theories \cite{esposito2005},
the best motivated alternatives to GR.
This is because for such systems, tensor-scalar theories predict the emission of a large 
amount of dipolar scalar waves in the form of gravitational radiation loss (as opposed to 
predominantly quadrupolar radiation predicted by GR).
This effect is strongly suppressed in double neutron star binaries, and as a result our timing 
measurements of PSR J1141$-$6545 impose much tighter constraints on these theories
than those possible by 
PSRs B1913+16, B1534+12 and J0737$-$3039 \cite{stairs2003,esposito2005}.
In fact, the measurement of an orbital decay that is in accord with GR would 
naturally exclude a significant part of the theory space.

A parameter of particular interest is the linear coupling strength \alphazero, for which the asymptotic 
limit scales 
as the square root of the measurement uncertainty in 
\Pbdot \cite{esposito2005}. 
This limit corresponds to 
strongly non-linear coupling 
(the quadrature coupling strength, \betazero $\sim$10 or larger). 
A simple extrapolation based on our current measurements yields 
$ \alphazerosq < 3.4 \times 10^{-6} $ in this regime. 
This is three times smaller than the Cassini bound ($ \alphazerosq \le 1.15 \times 10^{-5} $),
and thus provides the current best limit in that part of the theory space. 
For smaller values of \betazero (i.e. weakly non-linear coupling) however, the limit is weaker and is a 
function of \betazero.
In the special case of linear coupling (Brans-Dicke; \betazero = 0), 
it is reduced by a factor of 4\snsq, where \sns~is the self-gravity 
of the neutron star \cite{esposito2005}. 
Assuming 
\sns = 0.2, we derive $ \alphazerosq < 2.1 \times 10^{-5} $ for this 
regime, which is nearly twice the Cassini limit. 
Thus, the limit imposed by our current data
is weaker than that from Cassini for linear or 
weakly non-linear coupling but becomes increasingly stringent for strongly non-linear coupling 
and it should improve upon the Cassini bound somewhere within the range 5 \la \betazero \la 10. 
As our timing precision improves over time, these limits are expected to become even more stringent,
eventually cutting well below the Cassini bound at much lower values of \betazero.

\medskip

{\it Conclusions and Future Prospects.}---Our long 
timing campaign on PSR J1141$-$6545
has led to precise measurements of its PK parameters \omegadot, 
$\gamma$, and \Pbdot, providing strong confirmation of GR in gravitationally asymmetric 
binary systems. 
The measured orbital decay is in agreement with the GR prediction and the Shapiro delay is 
consistent with independent measurements of the orbital inclination angle. 
The pulsar and companion masses are determined to be \Mp = $1.27 \pm 0.01$ \Msun and 
\Mc = $1.02 \pm 0.01$ \Msun respectively. Stringent limits are placed on tensor-scalar theories of 
gravity, with an asymptotic limit of $ \alphazerosq < 3.4 \times 10^{-6} $, cutting well below the 
Cassini limit for large values of  the coupling parameter 
\betazero and hence providing the lowest 
limit in that range. 
For small values of \betazero, the limit is $ \alphazerosq < 2.1 \times 10^{-5} $ and 
thus weaker than that derived from the Cassini experiment.

Continued timing of this pulsar looks very promising. 
Currently \Pbdot is determined at a 6\% precision, the uncertainty in which is expected to decrease 
as $T^{-5/2}$, where $T$ is the observing time span. 
We thus anticipate approaching a precision near 2\% by 2012.
At such high precision, contaminations from the kinematic and Galactic contributions will start 
dominating the error budget \cite{dt91}.
While the Galactic term \PbdotGal is hard to determine accurately, 
the kinematic Shklovskii contribution \Pbdotkin can be 
assessed better when an independent measurement becomes available for the pulsar's transverse 
motion. 
Fortunately, the pulsar's location is such that \PbdotGal (dominated by differential rotation 
in the plane of the Galaxy) is of opposite sign to \Pbdotkin, and thus may cancel out a large 
fraction of it if the pulsar is moving at $\sim$100 \velu or faster, potentially 
resulting in a net contamination that is well below 1\%.
This will enable even more precise tests in the future and will place the most stringent constraints 
on alternative theories of gravity.

{\it Acknowledgements:} The Parkes Observatory is part of the Australia Telescope, funded by the 
Commonwealth of Australia for operation as a National Facility managed by CSIRO.
We are very grateful to Gilles Esposito-Far{\`e}se and Thibault Damour for several insightful discussions
and for a critical reading of the manuscript and to Willem van Straten for assistance with observations 
and fruitful discussions on timing analysis.

\bibliography{1141physrev}


\end{document}